\documentstyle[prd,aps,preprint,epsfig,floats]{revtex}
\tightenlines

\input epsf.tex
\def\DESepsf(#1 width #2){\epsfxsize=#2 \epsfbox{#1}}
\begin{document}

\draft

%@twocolumnfalse\endcsname
\preprint{\vbox{
\hbox{\bf UMD-PP-03-68}
}}

\title{{\Large\bf  Minimal SUSY SO(10) Model and Predictions for
Neutrino Mixings and Leptonic CP Violation}}

\author{\bf H. S. Goh, R.N. Mohapatra and Siew-Phang Ng }

\address{ Department of Physics, University of Maryland, College Park, MD
20742, USA}
\date{August, 2003}
\maketitle

\begin{abstract}
We discuss a minimal Supersymmetric SO(10) model where B-L symmetry is
broken by a {\bf 126} dimensional Higgs multiplet which also contributes
to fermion masses in conjunction with a {\bf 10} dimensional superfield.
This minimal Higgs choice provides a partial unification of neutrino
flavor structure with that of quarks and has been shown to predict all
three neutrino mixing angles and the solar mass splitting in agreement
with observations, provided one uses the type II seesaw formula for
neutrino masses. In this paper we generalize this analysis to include
 arbitrary CP phases in couplings and vevs.  We find that (i) the
predictions for neutrino mixings are similar with $U_{e3}\simeq 0.18$ as
before and other parameters in a somewhat bigger range and (ii) that to
first order in the quark mixing parameter $\lambda$ (the Cabibbo angle),
the leptonic mixing matrix is CP conserving. We also find that in the
absence of any higher dimensional contributions to fermion masses, the CKM
phase is different from that of the standard model implying that there
must be new contributions to quark CP violation from the supersymmetry
breaking sector. Inclusion of higher dimensional terms however allows the
standard model CKM phase to be maintained.
 \end{abstract}

\section{Introduction}
The observation of solar and atmospheric neutrino deficits in various
experiments such as Chlorine,
Super-Kamiokande, Gallex, SAGE and SNO together with
recent results from the K2K and Kamland\cite{solar}
experiments that involve terrestrial neutrinos have now conclusively
established that neutrinos have mass and they mix among themselves.
In conjunction with negative results from CHOOZ and PALO-VERDE reactor
experiments, one now seems to have
 a clear idea about the mixing pattern among the three
generations of neutrinos. Of the three angles needed to characterize these
mixings i.e. $\theta_{12}$, $\theta_{23}$ and
$\theta_{13}$, the first two are large and the third is
small\cite{review}.

 As far as the mass pattern is concerned, there are three distinct
possibilities:
 (i) ``normal hierarchy'' where $m_1\ll m_2 \ll m_3 \simeq \sqrt{\Delta
m^2_{A}}$;
(ii) inverted one where $m_1\simeq m_2 \simeq \sqrt{\Delta m^2_{A}}\gg
m_3$;
(iii) quasi-degenerate, where $m_1\simeq m_2\simeq m_3$.

 The normal hierarchy is similar to
what is observed in the quark sector although the degree of hierarchy
for neutrinos is much less pronounced. Understanding this difference is
somewhat of
a challenge although grand unified theories generally imply similar
hierarchies for quarks as well as neutrinos and therefore provide a
partial explanation of this challenge.

 The observed mixing pattern for leptons however, is totally different
from what
is observed for quarks, posing a much more severe theoretical challenge,
in particular for  grand unified theories that unify quarks and leptons.
Both these questions have been the subject of many papers\cite{rev} that
propose different approaches to solve these problems.
 Our goal in this paper is to address them in the context
of a minimal SO(10) model, which provides an interesting way to resolve
both the mass and mixing problems without making any extra assumptions,
other than what is needed to derive the supersymmetric standard model
from SO(10).

The first question one may ask is: why SO(10) ? The answer is that we seek
a theory of neutrino mixings that must be part of a larger framework
that explains other puzzles of the standard model such as the origin
of matter, the gauge hierarchy problem, dark matter etc. As is explained
below, supersymmetric SO(10) has the right features to satisfy all the
above requirements.

Supersymmetric extension of the
standard model (MSSM) seems to be a very natural framework to
solve both the gauge hierarchy as well as the dark matter problems.
The nonrenormalization theorem for the superpotential solves an important
aspect of the gauge gauge hierarchy problem i.e. the
radiative corrections do not destabilize the weak scale. As far as the
dark matter puzzle goes, if one adds the extra
assumption that R-parity defined as $R=(-1)^{3(B-L)+2S}$ is also a good
symmetry of MSSM, then the
lightest supersymmetric particle, the neutralino is stable and can play
the role of dark matter. In fact currently all dark matter search
experiments are geared to finding the neutralino. An added bonus of
MSSM is that it also provides a nice way to
understand the origin of electroweak symmetry breaking.

Turning to neutrinos, the seesaw mechanism\cite{seesaw} which is the
simplest way to understand small neutrino masses, requires a
right handed neutrino with a large Majorana mass $M_R$.
Present data require that $M_R$ has to be very high and yet much smaller
than the Planck mass. This would suggest that perhaps there is a new symmetry
whose breaking is responsible for $M_R$; since symmetry breaking
scales do not receive destabilizing radiative corrections in
supersymmetric theories, one has a plausible explanation of why
$M_R\ll M_{P\ell}$. One such symmetry
is local B-L symmetry\cite{marshak}, whose breaking
scale $v_{BL}$ could be responsible for $M_R$ (i.e. $M_R=fv_{BL}$) since
the righthanded
neutrino has nonzero B-L charge and is massless in the symmetry limit.
The formula for neutrino
masses in the seesaw model is given by $m_\nu\sim -\frac{m^2_D}{fv_{BL}}$;
therefore their smallness is now very easily understood as a result of
$m_D\ll
v_{BL}$\footnote{The symmetry could also be a global symmetry without
conflicting with what is known at low energies\cite{cmp}; but we do not
discuss it here.}.
The smallest grand unification group that
incorporates both the features required for
the seesaw mechanism i.e. a righthanded neutrino and the local B-L
symmetry is SO(10).

Secondly, the seesaw mechanism also goes well with the idea of grand
unification. It is well known that if MSSM remains the only theory till a
very high scale, then the three gauge couplings measured at LEP and SLC
at the
weak scale can unify to a single coupling around $ M_U\simeq 2\times
10^{16}$ GeV. Curiously enough,
naive estimates for the seesaw scale needed to understand the atmospheric
neutrino oscillation within a quark-lepton unified picture implies that
 the seesaw scale must be around $v_{BL}\geq 10^{15}$ GeV.
Close proximity of $v_{BL}$ and $M_U$ suggests that
one should seek an understanding of the neutrino masses and mixings within
a supersymmetric grand unified theory.
While this already  makes the case for SUSY SO(10)\cite{so10} as a
candidate theory for neutrino masses quite a strong one, there is an
additional feature that make the case even stronger.

This has to do with understanding a stable dark matter naturally in MSSM.
From the definition of R-parity given above,
it is clear that in the limit of exact SO(10) symmetry, R-parity is
conserved\cite{rabi}. However, since ultimately, SO(10) must break down to
MSSM, one
has to investigate whether in this process R-parity remains intact or
breaks down. This is intimately connected with how the
right handed neutrino masses arise (or how local B-L
subgroup of SO(10) is broken). It turns out that the B-L symmetry needs to
be broken by {\bf 126} Higgs representation\cite{babu,lee,goran} rather
than {\bf 16} Higgses\cite{many} as explained below.

 It was pointed out in \cite{babu} that SO(10) with one {\bf 126} and
one {\bf 10} is very predictive in the neutrino sector without any
extra assumptions, while at the same time correcting a bad mass relation
among charged fermions predicted by the minimal
SU(5) model, i.e. $m_\mu(M_U)=m_s(M_U)$. There is a partial  unification
of flavor structure between the neutrino and quark sectors and since all
Yukawa coupling parameters of the model are determined by charged fermion
masses and mixings, there are no free parameters (besides an
overall scale) in the neutrino mass matrix. It is therefore a
priori not at all clear that the neutrino parameters predicted by
the model would agree with observations. In fact the initial
analysis of neutrino mixings in this model that
 used the type I seesaw formula and did not include CP phases did predict
neutrino mixings that are now in disagreement with data. In
subsequent papers\cite{lav,lee,brahma,takasugi,fukuyama}, this
idea was analyzed ( in some cases by including more than one {\bf
10} Higgses) to see how close one can come to observations. The
conclusion now appears to be that one needs CP violating phases to
achieve this goal\cite{fukuyama}\footnote{For a different class of
SO(10) models with {\bf 126} Higgs fields see \cite{chen}}.

Another approach is to use the type II seesaw formula for neutrino masses
\cite{seesaw2} i.e.
\begin{eqnarray}
{\cal M}_\nu \simeq f\frac{v^2_{wk}}{\lambda
v_{B-L}}-\frac{m^2_D}{fv_{B-L}}.
\end{eqnarray}
 In models which have asymptotic parity
symmetry such as left-right or SO(10) models, it is the type II seesaw
that is more generic.

A very interesting point about this approach, noted recently\cite{bajc},
is that use of the type II seesaw formula for the two generation subsector
of $\nu_\mu$ and $\nu_\tau$, and dominance of
the first term leads to a very natural understanding of maximal
atmospheric mixing  angle due to $b-\tau$ mass convergence at high scale.

Whether the above idea does indeed lead to a realistic picture for
all three neutrino generations was left unanswered in
ref.\cite{bajc}. To answer this question, a complete three
generation analysis of this model was carried out in \cite{goh}
where it was pointed out that the same $b-\tau$ convergence
condition that led to large atmospheric mixing angle also leads to
a large solar angle and also a small $\theta_{13}$. Furthermore,
it also resolves the mass puzzle for neutrinos (i.e. a milder
hierarchy for neutrinos than that for quarks) since it predicts
that $\sqrt{\Delta m^2_{\odot}/\Delta m^2_{A}}\simeq \lambda\simeq
0.22$, where $\lambda $ is the small quark mixing parameter in the
Wolfenstein parametrization of the CKM matrix. As there are no
free Yukawa coupling parameters in the neutrino sector, it is
quite amazing that all the neutrino parameters can come out
 in the right range.

In the analysis of  ref.\cite{goh}, CP violating phases were set to zero.
It was implicitly assumed that all known CP violating processes in this
model would arise from the supersymmetry breaking
sector which would then make the model completely realistic even though it
does not have the conventional CKM CP violation. CP violation is however a
fundamental problem in particle physics and its origin at the moment is
unclear. It is therefore of interest to see (i) whether the minimal SUSY
SO(10)  can remain predictive for neutrinos even after the parameters in
the model are allowed to become complex, thereby ushering in CP violation
into the quark sector in a direct way
and (ii) if any other useful information on the nature of CP violation
for both quarks and leptons can be gained in this model.

To study CP violation in this model, we generalize our earlier analysis
to make all Yukawa coupling
parameters as well as the vacuum expectation values (vev) complex.
One might suspect that since this will bring in several new
parameters, the model may lose its predictivity.  We find that
this is not so. We have seven phases, one of which can play the
role of the CKM phase. Due to the presence of a sum rule involving
the charged lepton and quark masses, despite the presence of extra
phases, the model still remains predictive and pretty much leads
to the same predictions with minor changes as in the case without
phase.

Assuming that b-tau mass convergence leads to maximal neutrino
mixing, constrains three of the phases to be equal and matching
the electron mass fixes two others. The remaining arbitrary phase is
associated with the up quark, whose tiny mass keeps this phase well hidden
from this discussion. The CKM phase however
turns out to be outside the one $\sigma$ region of the present central
value in the
 standard model fits\cite{nir}. That means that one will need
some contributions to the observed CP violation from the SUSY
breaking sector. We then observe that if we include the higher
dimensional contributions to the fermion masses which were ignored
before, only for the first generation (which can be done naturally
using an R-symmetry), this introduces only one new parameter which
relaxes the electron mass constraint but does not affect the
neutrino sector. We can maintain a CKM phase equal to that given
by the standard model fit (of about $60^0$).

An important outcome of this analysis is that in both cases, to first
order in Cabibbo angle $\lambda$, the leptonic mixing matrix is
CP conserving, which can therefore be used to test the model.

In our opinion, these observations have lifted the minimal SO(10)
with {\bf 126} to a realistic grand unification model for all
forces and matter and ought therefore be considered as a serious
candidate for physics beyond the standard model up to the scale of
$10^{16}$ GeV. Just like the minimal SUSY SU(5) grand unified
theory could be tested in proton decay searches, this minimal
version of SO(10) can be tested by future neutrino experiments.
Important experiments for this purpose are the planned long base
line experiments which will provide a high precision measurement
of the mixing parameter $\theta_{13}$ (also called $U_{e3}$) to
the level of $0.1$ or less.

This paper is organized as follows: we start in sec. 2 with a few
introductory remarks about minimal SO(10) with {\bf 126} Higgses; we then
sketch a derivation of the type II seesaw formula in sec. 3 and show
how $b-\tau$ convergence leads to an understanding of large neutrino
mixings; in sec
4, we discuss the minimal SO(10) model with all couplings and vevs real
 and discuss the prediction for neutrinos. Section 4 gives more
details of our earlier paper\cite{goh} and gives an analytic
explanation of the constraints on the model parameters. In sec. 5, we
discuss the effects of including the most general form of CP violating
phases and present our predictions for neutrino masses and
mixings without the higher dimensional operators; in this section, we also
comment on the implication of including higher dimensional
operators. In sec. 7, we summarize our results and present our
conclusions.

\section{A minimal SO(10) model}
In any SO(10) model, one needs several multiplets to break the
symmetry down to $SU(3)_c\times U(1)_{em}$. Usually, to break the
SO(10) group down to the Pati-Salam group $SU(2)_L\times
SU(2)_R\times SU(4)_c$ or to the left-right group $SU(2)_L\times
SU(2)_R\times U(1)_{B-L}\times SU(3)_c$, one needs either {\bf 54}
or {\bf 54}$\oplus ${\bf 45} Higgs multiplets\cite{lee}. Then to
break $SU(2)_R\times U(1)_{B-L}$ down to $U(1)_Y$, one can employ
either ${\bf 16}+\overline{\bf 16}$ or  ${\bf 126}+\overline{\bf
126}$ pair since both these models have standard model singlet and
B-L$\neq$0 fields in them. The reason for having the complex
conjugate is to maintain supersymmetry down to the electroweak
scale. Finally to break the standard model group, {\bf 10 } is
used. Thus one needs always at least five Higgs multiplets in most
constructions of SO(10) model. One could replace {\bf 54}+{\bf 45}
pair by a {\bf 210} representation\cite{lee1}, reducing the number
of multiplets required to four. The neutrino results that we
discuss are not affected by the choice of Higgses that affect the
breaking in the first stage but are crucially dependent on how one
implements the subsequent ones.

In our model, we will use ${\bf 126} +\overline{\bf 126}$ to break B-L
symmetry for the following reasons.

\subsection{R-parity and {\bf 16} vs. {\bf 126}}
As discussed in the introduction, an important argument in favor
of MSSM being the TeV scale theory is the possibility that the
lightest SUSY partner can play the role of dark matter. In fact a
lot of resources are being devoted to discover the supersymmetric
dark matter particle. For MSSM to provide such a dark matter
particle, it is important that it has R-parity conservation. The
MSSM by itself does not have R-parity and ad hoc symmetries are
stuck into the MSSM to guarantee the existence of stable dark
matter. SO(10) provides an interesting way to guarantee automatic
R-parity conservation without invoking any ad hoc symmetry as we
see below.

The crucial question is whether B-L subgroup is broken by
a (i) {\bf 16}-dimensional Higgs field\cite{many} or (ii) {\bf
126}-dimensional ones\cite{babu,lee,goran}.

In case (i), B-L symmetry is broken
by a Higgs field (the $\nu_R$-like Higgs field in {\bf 16}) which has
$B-L=1$. If one looks at higher dimensional contributions to the
superpotential of the form ${\bf 16}^3_m{\bf 16}_H$, in terms of the MSSM
superfields they have the form
$QLd^c\nu^c_H$, $u^cd^cd^c\nu^c_H$ etc where $Q,L,u^c,d^c,e^c$ are the
matter superfields of the MSSM. The $\nu^c_H$ field is the Higgs field in
{\bf 16} that breaks B-L via $<\tilde{\nu}^c_H>= v_{BL}$. After symmetry
breaking, these
nonrenormalizable couplings will induce R-parity breaking terms of MSSM
such as $QLd^c$ and $u^cd^cd^c$ etc with a slightly suppressed coupling
i.e. $v_{BL}/M_{P\ell}\simeq 10^{-2}-10^{-3}$. This suppression is not
enough to let the neutralino play the role of cold dark matter - not to
mention the fact that it leads to extremely rapid proton decay.

It is sometimes argued that the final theory from which this effective
SO(10) model emerges may have additional local $U(1)_X$ symmetries that
will
prevent these dangerous higher dimensional terms. To see what this
implies, one can imagine  a Higgs field $X$ which is charged under
$U(1)_X$
and has vev of order of the GUT scale. The leading order operator that
will keep the theory safe from proton decay problem has to involve an
operator of the form $u^cd^cd^c\left(\frac{X}{M_{P\ell}}\right)^5$ and
similarly for other operators. This means that the $U(1)_X$ charge of X
and the operator $u^cd^cd^c$ must be arranged in a very specific way,
which leads to another kind of naturalness problem.

On the other hand if B-L symmetry of SO(10) is broken by a {\bf 126} Higgs
field, as in case (ii), the $B-L\neq 0$ and standard model singlet field
that breaks B-L
has B-L=2. As a result after this Higgs field acquires a vev, a $Z_2$
subgroup of B-L still survives and it keeps R-parity as a good symmetry.
This has been established by a detailed analysis of the superpotential in
Ref.\cite{goran}.
This not only forbids dangerous baryon number violating terms but also
allows for the existence of a neutralino dark matter without the need for
any additional symmetry. We will therefore work with an SO(10) model where
where the only field that breaks B-L is in the {\bf 126}-dimensional
representation.

\subsection{Mass sumrules in minimal SO(10)}
A second advantage of using {\bf 126} multiplet instead of {\bf
16} is that it unifies the charged fermion Yukawa couplings with
the couplings that contribute to righthanded as well as lefthanded
neutrino masses, as long as we do not include nonrenormalizable
couplings in the superpotential. This can be seen as
follows\cite{babu,lee}: it is the set {\bf 10}+${\bf
\overline{126}}$ out of which the MSSM Higgs doublets emerge; the
later also contains the multiplets $(3,1,10)+(1,3,\overline{10})$
which are responsible for not only lefthanded but also the right
handed neutrino masses in the type II seesaw formula. We explain
this below. Therefore all fermion masses in the model are arising
from only two sets of $3\times 3$ Yukawa matrices one denoting the
{\bf 10} coupling and the other denoting ${\bf \overline{126}}$
coupling.

In view of the above remarks, the SO(10) model that we will work
with in this paper has the following features: It contains three
spinor {\bf 16}-dim. superfields that contain the matter fields
(denoted by $\psi_a$); two Higgs fields, one in the {\bf 126}-dim
representation (denoted by $\Delta$) that breaks the
$SU(2)_R\times U(1)_{B-L}$ symmetry down to $U(1)_Y$ and another
in the {\bf 10}-dim representation ($H$) that breaks the
$SU(2)_L\times U(1)_Y$ down to $U(1)_{em}$. The original SO(10)
model can be broken down to the left-right group $SU(2)_L\times
SU(2)_R\times U(1)_{B-L}$ by ${\bf 54}\oplus {\bf 45}$ Higgs
fields denoted by $S$ and $A$ respectively.

To see what this model implies for fermion masses, let us explain how the
MSSM doublets emerge and the consequent fermion mass sumrules they
lead to. As noted, the
{\bf 10} and $\overline{\bf 126}$ contain two
(2,2,1) and (2,2,15) submultiplets (under $SU(2)_L\times SU(2)_R\times
SU(4)_c$ subgroup of SO(10)). We denote the two pairs by $\phi_{u,d}$
and $\Delta_{u,d}$. At the GUT scale, by some doublet-triplet
splitting mechanism these two pairs reduce to the MSSM Higgs pair
$(H_u,H_d)$, which can be expressed in terms of the $\phi$ and $\Delta$ as
follows:
\begin{eqnarray}
H_u &=& \cos\alpha_u \phi_u + e^{i\gamma_u} \sin\alpha_u \Delta_u \\ \nonumber
H_d &=& \cos\alpha_d \phi_d + e^{i\gamma_d}\sin\alpha_d \Delta_d
\end{eqnarray}
The details of the doublet-triplet splitting mechanism that leads to the
above equation are not relevant for what follows and we do not discuss it
here. As in the case of MSSM, we will assume that the Higgs doublets
$H_{u,d}$ have the vevs $<H^0_u>=v \sin\beta$ and $<H^0_d>=v \cos\beta$.

In orders to discuss fermion masses in this model, we start with the
SO(10) invariant superpotential giving the Yukawa couplings of the {\bf
16} dimensional matter spinor $\psi_i$ (where $i,j$ denote generations)
with the Higgs fields $H_{10}\equiv
{\bf 10}$ and $\Delta\equiv {\bf \overline{126}}$\footnote{For
alternative neutrino mass models with {\bf 126} representations
see\cite{chen}.}. \begin{eqnarray}
{W}_Y &=&  h_{ij}\psi_i\psi_j H_{10} + f_{ij} \psi_i\psi_j\Delta
\end{eqnarray}
SO(10) invariance implies that $h$ and $f$ are symmetric matrices.
We ignore the effects coming from the higher dimensional operators,
as we mentioned earlier.

Below the B-L breaking (seesaw) scale, we can write the superpotential
terms for the charged fermion Yukawa couplings as:
\begin{eqnarray}
W_0 &=& h_u QH_uu^c + h_d QH_d d^c + h_eLH_d e^c + \mu H_uH_d
\end{eqnarray}
where
\begin{eqnarray}
h_u &=& h\cos\alpha_u + fe^{i\gamma_u}\sin\alpha_u\\ \nonumber
h_d &=& h\cos\alpha_d + fe^{i\gamma_d}\sin\alpha_d\\ \nonumber
h_e &=& h\cos\alpha_d -3 fe^{i\gamma_d}\sin\alpha_d
\end{eqnarray}
In general $\alpha_u\neq \alpha_d$ and this difference is responsible for
nonzero CKM mixing
angles. In terms of the GUT scale Yukawa couplings, one can write the
fermion mass matrices (defined as ${\cal L}_m~=~\bar{\psi}_LM\psi_R$) at
the seesaw scale as:
\begin{eqnarray}
M_u &=& \bar{h} + \bar{f} \\ \nonumber
M_d &=& \bar{h}r_1 + \bar{f}r_2 \\ \nonumber
M_e &=& \bar{h}r_1 -3r_2 \bar{f} \\ \nonumber
M_{\nu^D} &=& \bar{h} -3 \bar{f}
\label{sumrule}\end{eqnarray}
where
\begin{eqnarray}
\bar{h} &=& h^* \cos\alpha_u \sin\beta\\ \nonumber
\bar{f} &=& f^* e^{i\gamma_u} \sin\alpha_u \sin\beta\\ \nonumber
r_1 &=& \frac{\cos\alpha_d}{\cos\alpha_u}\cot\beta\\ \nonumber
r_2 &=& e^{i(\gamma_d-\gamma_u)}\frac{\sin\alpha_d}{\sin\alpha_u}\cot\beta
\end{eqnarray}
The mass sumrules in Eq. (\ref{sumrule}) provide the first
important ingredient in discussing the neutrino sector. In the
case without any phases in the Yukawa sector, they determine
completely the input parameters of the model.

 To see this let us note that
Eq. (\ref{sumrule}) leads to the following sumrule involving the
charged lepton, up and down quark masses:
\begin{equation}\label{main}
    k \tilde{M}_l=r \tilde{M}_d+\tilde{M}_u
\label{kr}\end{equation} where $k$ and $r$ are complex numbers
which are functions of the symmetry breaking parameters of the
model; the mass matrices $M_{u,d,l}$ are general symmetric complex
matrices. In the Eq. (\ref{kr}), tilde denotes the fact that we
have made the mass matrices dimensionless by dividing them by the
heaviest mass of the species i.e. up quark mass matrix by $m_t$,
down quark mass matrix by $m_b$ etc.

We now proceed to do the phase counting in the model. First we absorb the
phase
of $r$ by redefining $k$, so that $r$ becomes real. We then choose
a basis so that $M_d$ is diagonal and real. The $(u,c,t)$ basis is
appropriately defined so that
the weak current is diagonal in this basis and $M_u$ is still a general
complex symmetric matrix wherein the CKM mixing matrix is buried. We can
now write
$M_u~=~V^TM^{diag}_uV$ where $V$ is a general $3\times 3$
unitary matrix, which has three real rotation angles and six phases.
Three of these phases can be put into the diagonal elements of the down
quark mass matrix and two can be put into the three diagonal elements of
$M_u$ and one remaining phase is in the CKM matrix. The $V$ matrix can now
be parameterized
as $V~=~P_dU_{CKM}P_u$, where $P_{u,d}$ are diagonal unitary matrices.

\noindent{\it Input parameters in the model:}

\begin{itemize}

\item We have two parameters $(k,r)$ and six phases constrained
by the fact that they must reproduce the correct charged lepton masses
and lead to large neutrino mixings via $b-\tau$ mass convergence.

\item  The
other degrees of freedom arise from the fact that the quark masses and
mixings extrapolated to the GUT scale have uncertainties in them.

\item We have the freedom to change the sign of the quark and lepton
masses, which amount to to a redefinition of the fermion fields by a
$\gamma_5$ transformation.
 \end{itemize}

We use the above input parameters to get the correct
charged lepton masses and subsequently via the type II seesaw (explained
below) to predict the neutrino mixings. As it is apparent,
this is a highly nontrivial task and restricts the parameter
space of the theory i.e. values of quark and charged lepton masses at the
GUT scale, very strongly.

\section{Type II seesaw formula and maximal neutrino mixings}
In this section, we explain the type II seesaw formula that we use in
discussing neutrino mixings.
The familiar seesaw formula (type I seesaw) for small neutrino
masses\cite{seesaw} is given by
\begin{eqnarray}
{\cal M}_\nu &=& -M_{\nu^D}M^{-1}_{N_R}M^T_{\nu^D}
\end{eqnarray}
where $M_{N_R}= f v_{BL}$, $v_{BL}$ being the scale of local B-L
symmetry breaking. On the other hand, it was pointed out in 1980 that
in theories with asymptotic parity conservation, the seesaw formula has an
additional contribution\cite{seesaw2} i.e.
\begin{eqnarray}
{\cal M}_\nu \simeq fv_L
-M_{\nu^D}{(fv_{BL})}^{-1}M^T_{\nu^D}.
\end{eqnarray}
where $v_L~=~\frac{v^2_{wk}}{\lambda v_{BL}}$.
Note that the $f$ matrix is common to both terms.
 It is possible to find different regions of parameter
space where the first or the second term may dominate. For example, it was
shown in Ref.\cite{darwin} that when parity symmetry is broken at a much
higher scale than the $SU(2)_R$ symmetry (e.g. by breaking SO(10) via
{\bf 210} Higgs field), one recovers the type I seesaw
formula.

In this section, we would like to make two points: first, the origin of
the first term in the SO(10) theory under consideration and
secondly, conditions under which the first term may dominate the neutrino
mass.

In our model, the mass of the right handed neutrino comes from a
renormalizable term of the form $\nu^c\nu^c\Delta^0_R$ where
$\Delta^0_R$ is the neutral member of an $SU(2)_R$ triplet; parity
symmetry of the theory then implies that there must be a coupling in the
theory of the theory of the left handed neutrino of the form
$\nu\nu\Delta^0_L$, where $\Delta^0_L$ is the neutral member of the left
handed triplet. If the $\Delta^0_L$ has a vev, then we get type II seesaw
formula.

To show that in our model, $\Delta^0_L$ has a vev, let us look at the
gauge invariant Higgs field terms in the superpotential.
 First we note the decomposition of the ${\bf \overline{126}}$
under the group $SU(2)_L\times SU(2)_R\times SU(4)_c$:
\begin{eqnarray}
{\bf \overline{126}} &=& (1,1,6)\oplus (2,2,15)\oplus
(3,1,\overline{10}) \oplus (1,3,10)
\end{eqnarray}
The $SU(2)_L$ triplet that contributes to the type II seesaw formula is
contained in the multiplet $\Delta_L\equiv (3,1,\overline{10})$ and it
couples to the left
handed multiplet $\psi \equiv (2,1,4)$ of the {\bf 16} dimensional
SO(10) spinor that contains the matter fermions
i.e. $\psi_L\psi_L\Delta_L$. On the other hand the mass of the RH
neutrinos comes from the coupling of $\Delta_R\equiv (1,3, 10)$
submultiplet of ${\bf \overline{126}}$ to the right handed lepton
doublets.

The vev of the neutral member of $\Delta_R$ breaks the B-L
symmetry and gives mass to the RH neutrinos. This generates the
second term in the type II seesaw formula. To see how the
$\Delta^0_L$ vev arises, note that the general superpotential of
the model contains terms of type $\lambda_1{\bf 126}^2\cdot {\bf
54}$ and $\lambda_2{\bf 10}\cdot {\bf 10}\cdot {\bf 54}$. In the
Higgs potential, this generates a term (from $|F_{\bf 54}|^2$) of
the form ${\bf 10\cdot 10\cdot \overline{126}\cdot
\overline{126}}$. In this expression, there is a term of the form
$\phi(2,2,1)^2\Delta_L(3,1,10)\Delta_R(1,3,\overline{10})$ with a
coefficient $\lambda_1\lambda_2$. Furthermore, in the Higgs
potential, there is a mass term for $\Delta_L(3,1,10)$ of the form
$\mu^2_{\Delta}+\lambda_3 v^2_U$, where $v_U$ is the GUT scale. On
minimizing the potential, these two terms lead to a vev for the
$SU(2)_L$ triplet $\sigma_L\equiv <\Delta^0_L>\simeq
\frac{\lambda_1\lambda_2 v^2_{wk} v_{BL}}{\mu^2_{\Delta}+\lambda_3
v^2_U}$.

 It is now clear that if we choose $\lambda_3$ such that
$\mu^2_{\Delta}+\lambda_3 v^2_U \ll v^2_{BL}$, then the entries in
the second matrix in the type II seesaw formula can much smaller
than the first term. When this happens, then Eq. (6) can be used
to derive the sumrule
\begin{eqnarray}
{\cal M}^*_{\nu} &=& a(M_{\ell}-M_d)
\label{key}\end{eqnarray}
This equation is key to our discussion of the neutrino masses and mixings.

\subsection{Maximal neutrino mixings from type II seesaw}
Using Eq. (\ref{key}) in second and third generation sector, one
can understand the results of \cite{bajc} in a heuristic manner as
follows. The known hierarchical structure of quark and lepton
masses as well as the known small mixings for quarks suggest that
the matrices $M_{\ell,d}$ for the second and third generation have
the following pattern
\begin{eqnarray}
M_{\ell}~\approx~m_\tau\pmatrix{\lambda^2 &\lambda^2\cr
\lambda^2 & 1}\\ \nonumber
M_q ~\approx~m_b \left(\begin{array}{cc}\lambda^2 & \lambda^2\cr \lambda^2
& 1\end{array}\right)
\end{eqnarray}
where $\lambda \sim 0.22$ (the Cabibbo angle) as is required by
low energy observations. It is well known that in supersymmetric
theories, when low energy quark and lepton masses are extrapolated
to the GUT scale, one gets approximately that $m_b\simeq m_\tau$.
One then sees from the above sumrule for neutrino masses Eq.
(\ref{key}) that all entries for the neutrino mass matrix are of
same order $\lambda^2$ leading very naturally to the atmospheric
mixing angle to be large. Thus one has a natural understanding of
the large atmospheric neutrino mixing angle. No extra symmetries
are assumed for this purpose.

For this model to be a viable one for three generations, one must
show that the same $b-\tau$ mass convergence at GUT scale also
explains the large solar angle $\theta_{12}$ and a small
$\theta_{13}$. This has been demonstrated in a recent
paper\cite{goh}.

To see how this comes about, let us ignore the CP violating phases
and recall Eq. (\ref{kr}). Note that in the basis where the down
quark mass matrix is diagonal, all the quark mixing effects are
then in the up quark mass matrix i.e. $M_u ~=~ U^T_{CKM}M^d_u
U_{CKM}$. Using the Wolfenstein parametrization for quark mixings,
we can conclude that that we have
\begin{eqnarray}
M_{d}~\approx ~m_{b}\pmatrix{\lambda^4 & \lambda^5
&\lambda^3\cr \lambda^5 & \lambda^2& \lambda^2 \cr \lambda^3 & \lambda^2 &
1}
\end{eqnarray}
and $M_{\ell}$ and $M_d$ have roughly similar pattern due to the
sum rule \ref{kr}. In the above equation, the matrix elements are
supposed to give only the approximate order of magnitude. As we
extrapolate the quark masses to the GUT scale, due to the fact
that $m_b-m_\tau \approx m_{\tau}\lambda^2$ for some value of
$\tan\beta$, the neutrino mass matrix $M_\nu~=~c(M_d-M_\ell)$
takes roughly the form\cite{nussinov}
\begin{eqnarray}
M_{\nu}~=~c(M_d-M_\ell)\approx ~m_0\pmatrix{\lambda^4 & \lambda^5
&\lambda^3\cr \lambda^5 & \lambda^2 & \lambda^2 \cr \lambda^3 & \lambda^2
& \lambda^2}
\end{eqnarray}
 It is then easy to see from this mass matrix that both the $\theta_{12}$
(solar angle) and $\theta_{23}$ (the atmospheric angle) are
large. Furthermore, if only terms of $\lambda^2$ are kept, then,
$m_3\approx c\lambda^2$ and in the limit of maximal atmospheric mixing and
both $m_2$ and $m_1$ vanish. As soon as terms of order $\lambda^3$,
$m_{1,2}$ pick up mass and then one has $m_2/m_3\approx \lambda$. This
then naturally explains the milder hierarchy among neutrinos compared to
that among quarks.

The detailed predictions of the model such as the magnitudes of these
angles and neutrino masses $m_{1,2}$ depend on the details of the
quark masses at the GUT scale and we discuss it in the following
sections for different cases.

\section{Predictions for neutrino mixings without any CP phases in the
mass sumrules} Let us first consider the case where the Yukawa
couplings in the superpotential and the vevs of doublet Higgs are
all real. This case was considered in our earlier paper\cite{goh}.
In this case all CP phases needed for understanding the observed
kaon and B CP violation arise from the supersymmetry breaking
sector. We can start by solving for the parameters $k$ and $r$ in
Eq. (\ref{kr}) and find the range of quark masses for which the
charged lepton masses come out right. We then use the values of
$k$ and $r$ as well as the quark masses to get the neutrino masses
and mixings using Eq. (\ref{key}).

While all our predictions are done via detailed numerical analysis
using Mathematica, in this section we provide a qualitative
discussion of the nature of the constraints on the model parameters.
The qualitative discussion brings out several things clearly:

(i) while
the masses of tau lepton and muon fix the values of $k$ and $r$, getting
electron mass is nontrivial and requires fine tuning for quark masses and
also the $V_{ub}$ parameter within the range allowed by present data.

(ii) Secondly, we derive an approximate form for the neutrino mass matrix
and and show how the model generally tends to predict values of
$U_{e3}$ close to its present upper limit.

To find $k$ and $r$ numerically, we need to specify the values of the
quark and lepton masses as well as the CKM mixings extrapolated to the GUT
scale. These have been discussed extensively in literature. We use the
values from the paper of Das and Parida\cite{parida} and are given in
Table.

\bigskip
\begin{center}
\begin{tabular}{|c||c||c|}\hline
input observable & $\tan\beta=10$ & $\tan\beta=55$ \\ \hline $m_u$
(MeV) & $0.72^{+0.13}_{-0.14}$ & $0.72^{+0.12}_{-0.14}$\\\hline
$m_c$(MeV) & $210.32^{+19.00}_{-21.22}$ &
$210.50^{+15.10}_{-21.15}$ \\ \hline $m_t$(GeV) &
$82.43^{+30.26}_{-14.76}$ & $95.14^{+69.28}_{-20.65}$ \\\hline
$m_d$ (MeV) & $1.50^{+0.42}_{-0.23}$ &$1.49^{+0.41}_{-0.22}$
\\\hline
$m_s$ (MeV) & $29.94^{+4.30}_{-4.54}$ &$29.81^{+4.17}_{-4.49}$
\\\hline
$m_b$ (GeV) & $1.06^{+0.14}_{-0.08}$ &$1.41^{+0.48}_{-0.19}$
\\\hline
$m_e$ (MeV) & $0.3585$ & $0.3565$ \\\hline
$m_{\mu}$(MeV) & $75.6715^{+0.0578}_{-0.0501}$ &
$75.2938^{+1912}_{-0.0515}$\\\hline
$m_{\tau}$ (GeV) & $1.2922^{+0.0013}_{-0.0012}$ &
$1.6292^{+0.0443}_{-0.0294}$\\ \hline
\end{tabular}
\end{center}
\bigskip
Taking the 2-3 submatrix of the Eq. (\ref{kr}) and remembering
that all mixing angles are small, we get from this equation the
approximate equations:
\begin{eqnarray}
k & \simeq & r+1\\ \nonumber k\frac{m_\mu}{m_\tau}& \simeq &
r\frac{m_s}{m_b} + \frac{m_c}{m_t}
\end{eqnarray}
Using these equations and rough numbers for $m_\mu/m_\tau \sim 0.059$ and
$m_s/m_b\simeq -0.026$, we get,  $k\simeq 0.29 $ and $r\simeq -0.71$. The
results of our detailed numerical solutions for this case give
$-0.78 \leq r \leq -0.74$ and $0.23 \leq k \leq 0.26$.

Now we illustrate with a particular choice of fermion masses how
values of $k$ and $r$ are determined and then show how the small electron
mass comes out in the model. For
this purpose, let us express the $\tilde{M}_\ell$ in terms of the small
parameter $\lambda\sim 0.22$ (the Cabibbo angle) for a particular choice
of $m_b(M_U)\simeq 1.2$ and $m_\tau\simeq 1.29$ as an example:
\begin{eqnarray}
\tilde{M}_\ell &=& \pmatrix{\mp 1.6\lambda^4\pm 11\lambda^6 & \sim
6\lambda^5 &
4.5\lambda^3\cr 6\lambda^5 & \mp 2.1 \lambda^2 & 3.3\lambda^2 \cr  4.5
\lambda^3 & 3.3\lambda^2 & 1}
\end{eqnarray}
where we have kept the $\pm$ sign in $M_{\ell, 11}$ to reflect the sign
freedom in the quark masses.

An approximate expression for the electron mass $\tilde{m}_e$ can be
obtained from the above expression to be:
\begin{eqnarray}
\tilde{m}_e & \simeq & M_{\ell, 11} - \frac{M^2_{\ell, 13}}{M_{\ell, 33}}\\
\nonumber
& \simeq & \mp 1.6\lambda^4sgn(m_d) \pm sgn(m_c) 10\lambda^6
-20\lambda^6
\end{eqnarray}
Since $\tilde{m}_e\sim \lambda^5$, we need a cancellation between the
different terms in the above equation. Note that if $m_d< 0$ and $m_c <
0$, then we get cancellation between the three terms in the above equation
and we can get the right value for the electron mass.
 It must however be stressed
that $m_e$ is reproduced only for particular choice of the bottom and tau
masses at the GUT scale and of course these are in the allowed domain but
nonetheless they reflect the constraint on the mass parameters for the
model to be acceptable. It is impressive that it works.

We caution that the above discussion is meant to give a flavor of the
constraints on the model. In the detailed numerical calculations, the
range of parameters where correct electron mass results is larger than
what would be implied by the above discussion.

To study the neutrino masses and mixings, let us write down the neutrino
mass matrix in this model:
\begin{eqnarray} {\cal M}_\nu &=& \pmatrix{ y\tilde{m_d} -z
(\tilde{m_c}\lambda^2+A^2\lambda^6|\Lambda|^2 & \cdot\cdot\cdot &
\cdot\cdot \cr
z\tilde{m_c}\lambda+zA^2\lambda^5|\Lambda| & y\tilde{m_s}-z\tilde{m_c}
-zA^2\lambda^4 &\dot\cdot\cdot\cr
zA\tilde{m_c}\lambda^3+zA\lambda^3|\Lambda| & -zA\lambda^2(1+\tilde{m}_c)
& y-z\tilde{m_c}A^2\lambda^4- z }
\end{eqnarray}
where $4y~=~(m_b/m_\tau-r/k)$ and $4z~=~k^{-1}$ and $\Lambda =
(1-\rho-i\eta)$.
Noting that $\tilde{m}_d\simeq \lambda^4$, $\tilde{m}_c\simeq \lambda^4$,
the neutrino mass matrix to leading order takes the form
\begin{eqnarray}
{\cal M}_\nu~=~\pmatrix{0 & 0 & zA\lambda^3|\Lambda|\cr 0 & y\tilde{m}_s
&zA\lambda^2\cr zA\lambda^3|\Lambda| & zA\lambda^2 &y-z }
\end{eqnarray}
 From this we see that
 maximal neutrino mixing as well as the correct mass hierarchy comes out.
Obviously to get the mixing angles in the desired range, detailed analysis
is needed and we have carried it out.

Secondly, one can give an ``analytic'' argument that $U_{e3}$ will
be close to its present upper limit. Again Eq. (\ref{kr}) comes in
handy. Roughly $U_{e3}\simeq \frac{{\cal M}_{\nu, 13}}{{\cal
M}_{\nu, 33}-{\cal M}_{\nu, 11}}\simeq
\frac{m_\tau\lambda^3}{m_b(M_U)-m_\tau(M_U)}\simeq \lambda$.

\subsection{Detailed Numerical Analysis}
We undertake extensive scanning of the parameter space of the model
defined by the uncertainties in the values of the fermion masses at the
GUT scale and the values of $k$ and $r$ in the neighborhood of the values
given above. We have used the
values of the standard model fermion masses from Table I and the following
values for the mixing angles.
\begin{eqnarray}
U_{CKM} &=& \pmatrix{0.974836& 0.222899& -0.00319129\cr -0.222638 &
0.974217 &  0.0365224\cr 0.0112498& -0.0348928& 0.999328}
\end{eqnarray}
The strategy in our numerical calculations is the following: we
focus on the Eq. (\ref{kr}) and using the inputs in the right hand
side, we look for the eigenvalues of the charged lepton mass
matrix in the left hand side to match the observed lepton masses.
Since there are only two unknowns, the second and third generation
masses largely fix the parameters $k,r$ as we just described. To
match the electron mass is nontrivial. The parameters are
essentially the signs of the fermion masses and the present
uncertainties in the values of the SM fermion masses. After we fix
the electron mass it narrows down our parameter range somewhat. We
then look for neutrino masses and mixing angles using Eq.
(\ref{key}) in the remaining (small) parameter range. Note that
due to overall scale freedom of the type II seesaw scale, we
cannot predict the $\Delta m^2_{A}$.  We also do a direct
numerical solution. Both the results are in agreement.

 The solutions we present
here correspond to $m_{e,\mu,\tau,b,t} > 0$ and $m_{c,d,s}< 0$ up to
an overall sign.

Our results are displayed in Fig. 1-3 for the case of the
supersymmetry parameter $\tan\beta~=~10$. In these figures, we
have restricted ourselves to the range of quark masses for which
the atmospheric mixing angle $\sin^22\theta_A \geq 0.8$. (For
presently preferred range of values of $\sin^22\theta_A$ from
experiments, see \cite{kan}).
 We then present the predictions for
$\sin^22\theta_{\odot}$, $\Delta m^2_{\odot}$ and $U_{e3}$ for the allowed
range $\sin^22\theta_A$ in Fig.1, 2 and 3 respectively.
 The spread in the predictions come
from uncertainties in the $s,c$ and the $b$-quark masses. Note two
important predictions: (i) $\sin^22\theta_{\odot}\geq 0.91$ and $U_{e3}\sim
\pm 0.16$. The present allowed range for the solar mixing angle is
$0.7 \leq \sin^22\theta_{\odot} \leq 0.99$ at 3$\sigma$
level\cite{kan,bahcall}.
The solutions for the neutrino mixing angles are
sensitive to the $b$ quark mass.

It is important to note that this model predicts
the $U_{e3}$ value very close to the present experimentally allowed upper
limit and can therefore be tested in the planned long base line
experiments which are expected to probe
$U_{e3}$ down to the level of $\sim 0.05$\cite{numi,JHF}.

\begin{figure}[h!]
\begin{center}
\epsfxsize15cm\epsffile{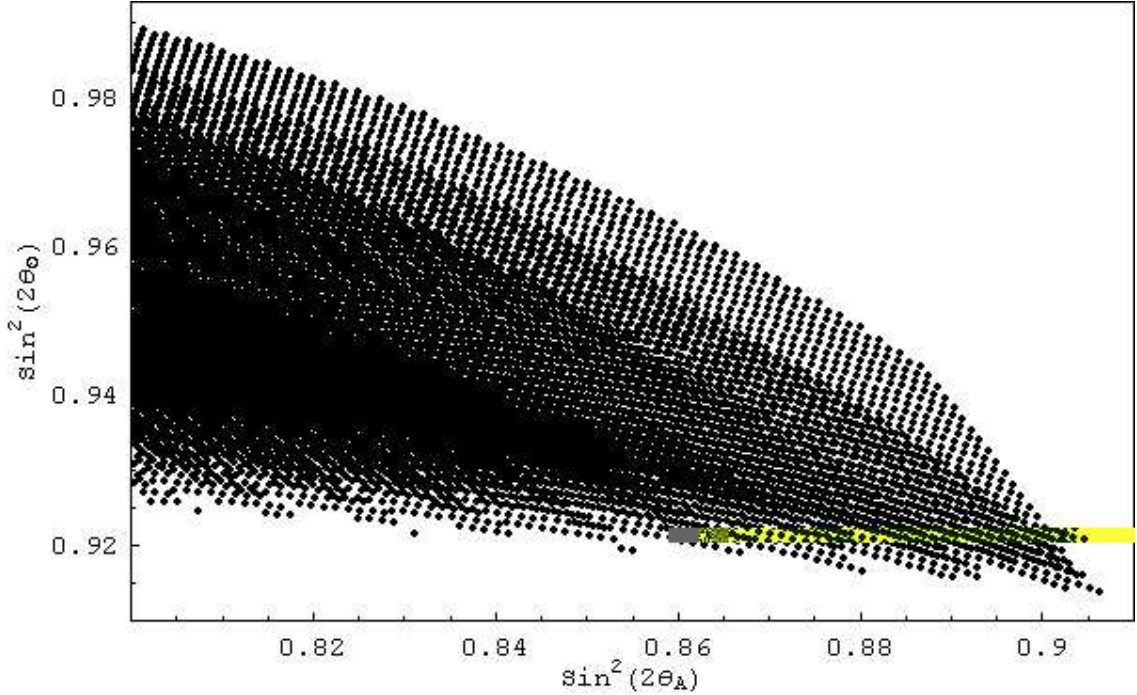}
\caption{
The figure shows the range of predictions for $\sin^22\theta_{\odot}$ and
$\sin^22\theta_A$ for the range of quark masses in table I that fit
the charged lepton spectrum and where all CP phases are set to zero. We
required that the ratio $\Delta
m^2_{\odot}/\Delta m^2_A \leq 0.05$. Note
that $\sin^22\theta_{\odot}\geq 0.9$ and $\sin^22\theta_A\leq 0.9$.
\label{fig:cstr1}}
\end{center}
\end{figure}

\begin{figure}[h!]
\begin{center}
\epsfxsize15cm\epsffile{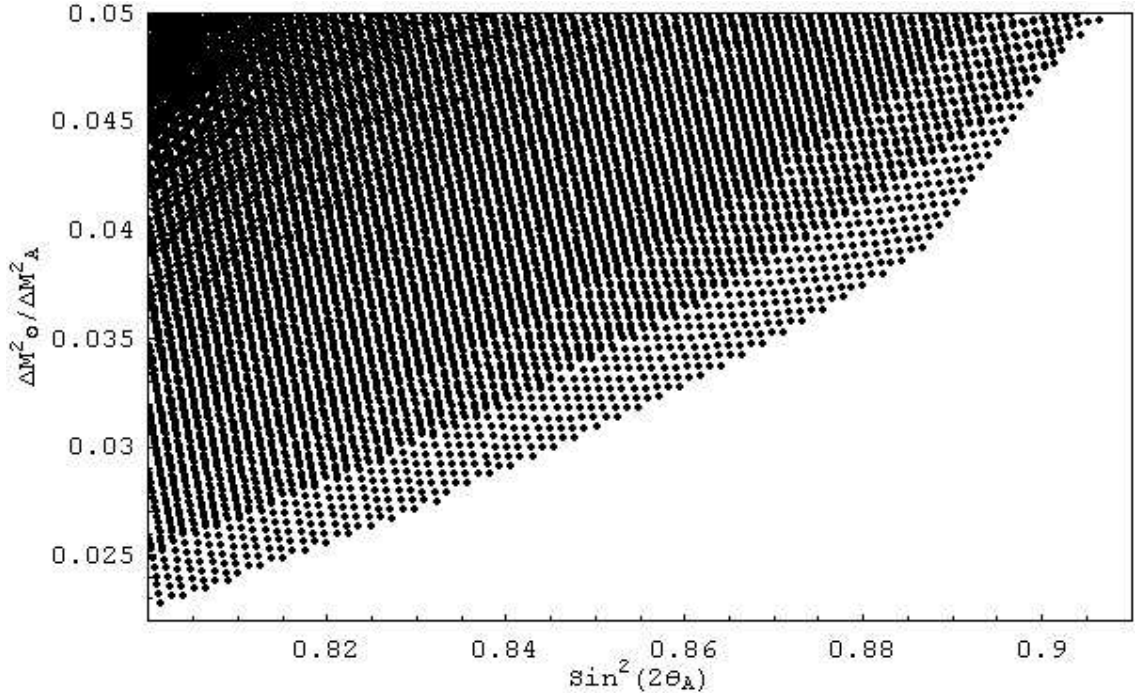}
\caption{
The figure shows the predictions for $\sin^22\theta_{A}$ and
 $\Delta m^2_{\odot}/\Delta m^2_{A}$ for the range of quark masses and
mixings that fit charged lepton masses and where all CP phases in the
fermion masses are set to zero.
\label{fig:cstr2}}
\end{center}
\end{figure}

\begin{figure}[h!]
\begin{center}
\epsfxsize15cm\epsffile{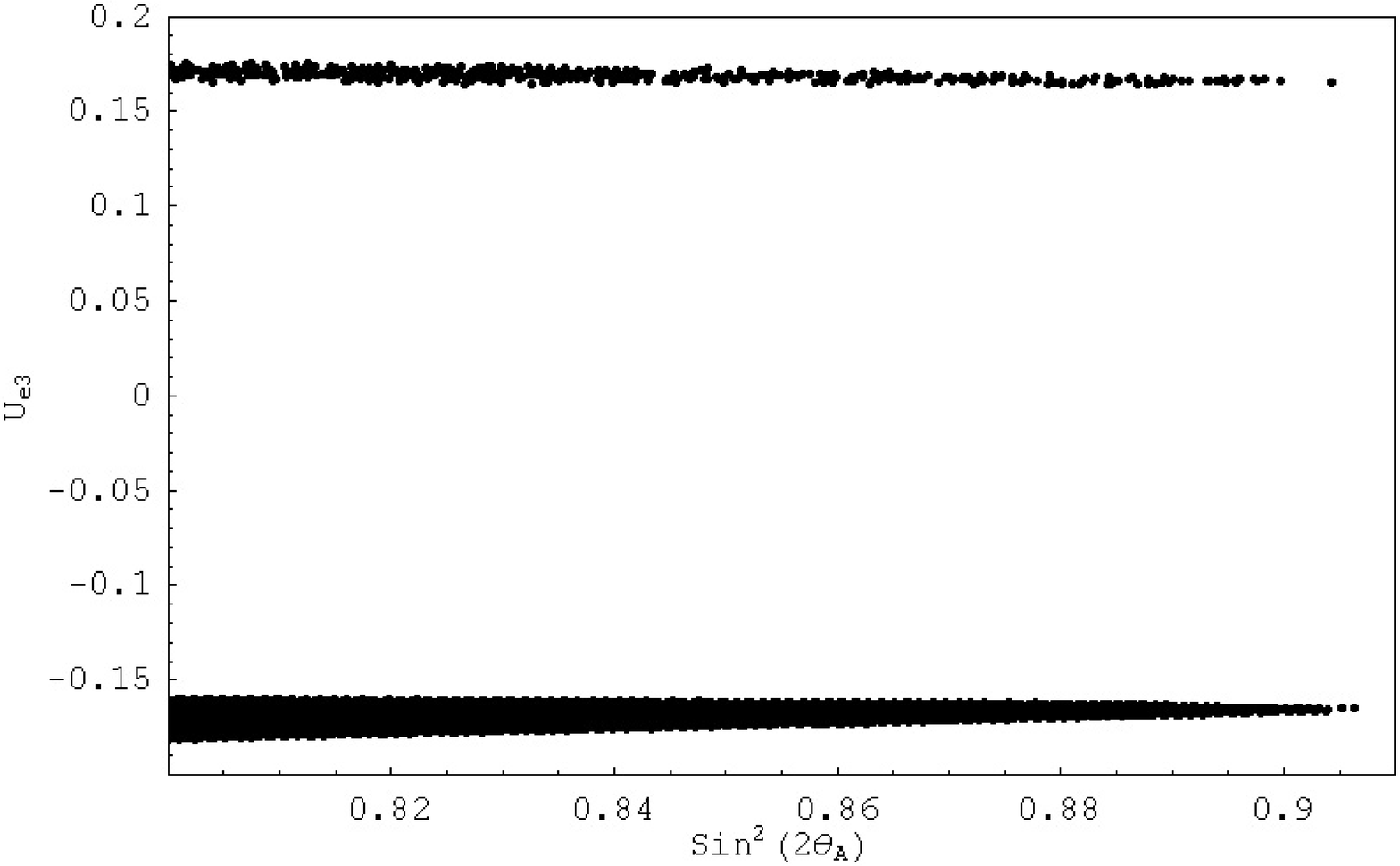}
\caption{
The figure shows the predictions of the model for $\sin^22\theta_{A}$
and $U_{e3}$ for the allowed range of parameters in the model as in
Fig. 1 and 2. Note that
$U_{e3}$ is very close to the upper limit allowed by the existing reactor
experiments.
\label{fig:cstr3}}
\end{center}
\end{figure}

 \section{Effect of CP phases on neutrino mixings}
In this section, we keep the CP phases as described before and
look for solutions to the charged lepton equation, Eq. (\ref{kr})
and then use the allowed parameter range to look at the
predictions of the neutrino masses and mixings. We have seven CP
phases including the CKM phase.
 One might think that since there
are more parameters in this case, getting a solution will be trivial. We
find that actually, it is not so. Let us demonstrate this in an analytic
way very crudely and we follow it up with detailed numerical scan to get
the mixing angles.

To proceed with this analysis, first note that the phases are
distributed in Eq. (\ref{kr}) as follows. First the down quark
mass eigenvalues are chosen to be complex i.e.
$(\tilde{m_d}e^{i\beta_1}, \tilde{m_s}e^{i\beta_2}, e^{i\beta_3})$
and similarly for the up quark sector, there are three phases in
the masses (denoted by $\alpha_i$).
 Using \ref{kr}, we
see that

\begin{eqnarray} k M_\ell &=& \pmatrix{ r\tilde{m_d}e^{i\beta_1}
+\tilde{m_c}\lambda^2e^{i\alpha_2} &\tilde{m_c}\lambda
e^{i\alpha_2}+A^2\lambda^5e^{i\alpha_3}(\Lambda) &A\tilde{m_c}\lambda^3
e^{i\alpha_2}+A\lambda^3e^{i\alpha_3}(\Lambda)
\cr
+A^2\lambda^6e^{i\alpha_3}(\Lambda)^2 & & \cr
\tilde{m_c}\lambda
e^{i\alpha_2}+A^2\lambda^5e^{i\alpha_3}(\Lambda)   &
r\tilde{m_s}e^{i\beta_2}+\tilde{m_c}
e^{i\alpha_2}+A^2\lambda^4e^{i\alpha_3} & \tilde{m_c}A\lambda^2
e^{i\alpha_2}+A\lambda^2e^{i\alpha_3} \cr
 & & \cr
A\tilde{m_c}\lambda^3
e^{i\alpha_2}+A\lambda^3e^{i\alpha_3}(\Lambda)   &
\tilde{m_c}A\lambda^2 e^{i\alpha_2}+A\lambda^2e^{i\alpha_3}   &
re^{i\beta_3}+\tilde{m_c}A^2\lambda^4 e^{i\alpha_2}+e^{i\alpha_3}
}
\end{eqnarray}
where we have used the Wolfenstein parametrization for the CKM matrix.
Essentially the eigenvalues of the matrix on the right hand side must
match the charged lepton masses. Note that as in the case without the CP
phase, this will require us to work
only in a very limited range of the quark masses. Since now we have to
align the phases with the known standard model CKM phase, the
constraints are even tighter. The muon and tau masses
come out quite easily as in the case without phases discussed in the
previous section.

As in the CP conserving case, the mass of the electron requires fine
tuning as can be
seen by noting that the $11$ entry of $M_{\ell}$ is roughly of order
$\lambda^4$ which is about 3 to 4 times
the value required to give the correct electron mass. One must therefore
have cancellation to get the correct $m_e$ .  Before discussing the
details of this cancellation, let us first look at the neutrino mass
matrix, which looks as follows:

\begin{eqnarray}
{\cal M}_\nu &=& \pmatrix{0 & 0 & zA\lambda^3 (1-\rho-i\eta)e^{i\alpha_3}\cr
0 & 0.7y\lambda^2e^{\beta_2} & zA\lambda^2e^{i\alpha_3} \cr
zA\lambda^3 (1-\rho-i\eta)e^{i\alpha_3} &zA\lambda^2e^{i\alpha_3} &
ye^{i\beta_3}-ze^{i\alpha_3}}
\end{eqnarray}
First thing to note is that we must have $\alpha_3 = \beta_3$ for
$b-\tau$ unification to lead to maximal mixing.

To see very qualitatively what value of CKM phase is required for the
model to work, we need
to analyze the charged lepton masses. As in the CP conserving case, the
muon and the tau mass come out for the choice of $k \approx 0.25$ and
$r\approx -.75$. Let us now discuss the electron mass.
For this we first note that we use the Wolfenstein parametrization. The
two parameters $\rho$ and $\eta$ responsible for the CKM phase are given
by the latest data to be\cite{nir} $.12 \leq \rho \leq .35$ and $0.28 \leq
\eta \leq .41$.
 Since the charged lepton mass matrix has hierarchical structure,
we can deduce an approximate formula for $m_e$ from the sum rule \ref{kr}
and using the above values of $k$ and $r$:
\begin{eqnarray}
\tilde{m}_e\simeq
-3\tilde{m_d}e^{i\beta_1}+\tilde{m_c}\lambda^2
e^{i\alpha_2}+A^2\lambda^6e^{i\alpha_3}(\Lambda)^2
-{(4A\lambda^6(1-\rho-i\eta)^2
e^{i\alpha_3})}- O(\lambda^7)
\end{eqnarray}
Numerically, to get the correct value for $\tilde{m}_e\simeq
0.00028\simeq 0.6 \lambda^5$, we must cancel most of the
$\tilde{m}_d$ contribution to it i.e. $-3\tilde{m}_d\simeq
1.3\lambda^4$, from the other terms: these terms get maximized for
$\rho < 0$ in which case we get $|1-\rho-i\eta|\simeq 1.41$ so
that the last term contributes $32\lambda^6\simeq 1.5 \lambda^4$.
It also determines the arbitrary phases that accompany the quark
masses as follows: $\beta_1=\alpha_2=\alpha_3+2\delta$ where
$\tan\delta\simeq \frac{\eta}{1- \rho}$ and $m_d < 0$. The terms
in the electron mass formula then cancel leading to $\tilde{m}_e
\simeq .9 \lambda^5$ which is in the rough ``neighborhood'' of the
observed
 value ($\tilde{m}_e$).
This value of $\rho$ gives a CKM phase which is in the third
quadrant implying that in order to understand the observed CP
violation i.e. $\sin 2\beta$ in B-system, one will need to invoke
CP violation from the supersymmetry breaking scalar masses.

Again we caution the reader that this is a crude estimate. In
detailed numerical analysis, we can get solutions for small
positive $\rho$ and one next has to see how these values fit the
neutrino mixings. Thus combined fit to both electron mass and
neutrino mixings works only for negative $\rho$ value which is
different from the current standard model CKM fit.

The detailed numerical predictions for this model are given in Fig. 4,5
and 6, where we have scanned over all the phases looking for the correct
charged lepton masses and acceptable neutrino parameters. We have also
extended the domain of $\Delta m^2_{\odot}/\Delta m^2_{A}$ to $0.08$.

\begin{figure}[h!]
\begin{center}
\epsfxsize15cm\epsffile{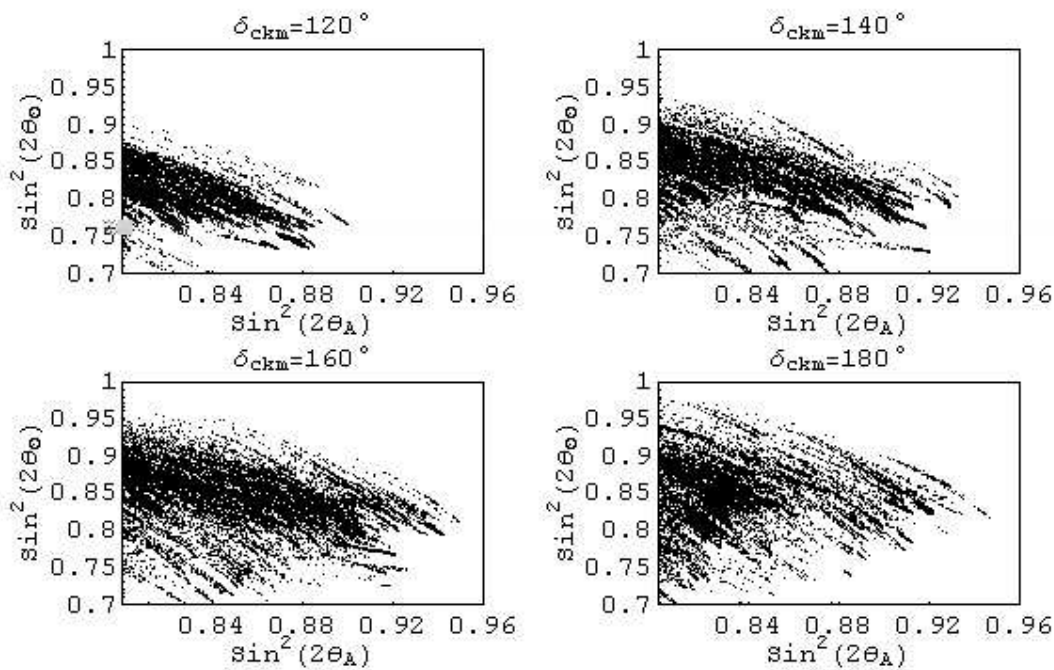}
\caption{
The figure shows the predictions for $\sin^22\theta_{A}$ and
 $\sin^22\theta_{\odot}$ for the range of quark masses and
mixings that fit charged lepton masses in the presence of all CP
phases in the fermion masses. The four different panels give the
predictions for different values of the CKM phase
($\tan\delta_{CKM}=\frac{\eta}{\rho}$). Note that all these values
are outside the one sigma region of the present standard model fit
to all CP violating data. Note that the case $\delta_{CKM}=180$
includes the CP conserving case discussed in sec. IV.  \label{fig:cstr4}}
\end{center}
\end{figure}

\begin{figure}[h!]
\begin{center}
\epsfxsize15cm\epsffile{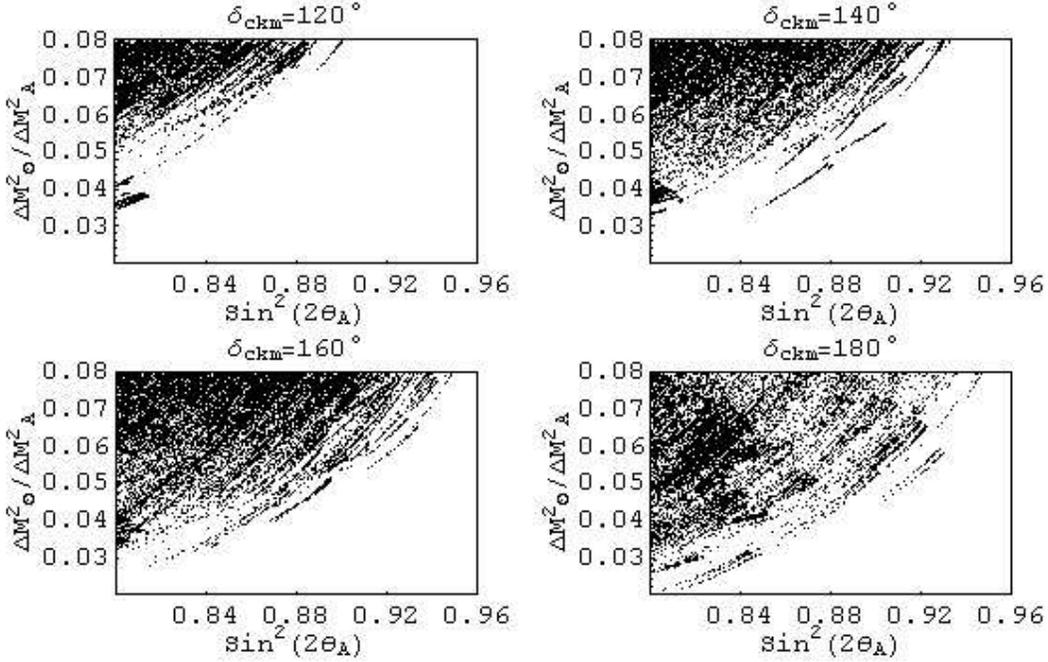}
\caption{
The figure shows the predictions for $\sin^22\theta_{A}$ and
 $\Delta m^2_{\odot}/\Delta m^2_{A}$ for the range of quark masses and
mixings that fit charged lepton masses and where all CP phases in the
fermion masses are kept subject to the condition that b-tau mass
convergence is responsible for large neutrino mixings, for four different
values of the $\delta_{CKM}$.
\label{fig:cstr5}}
\end{center}
\end{figure}

\begin{figure}[h!]
\begin{center}
\epsfxsize15cm\epsffile{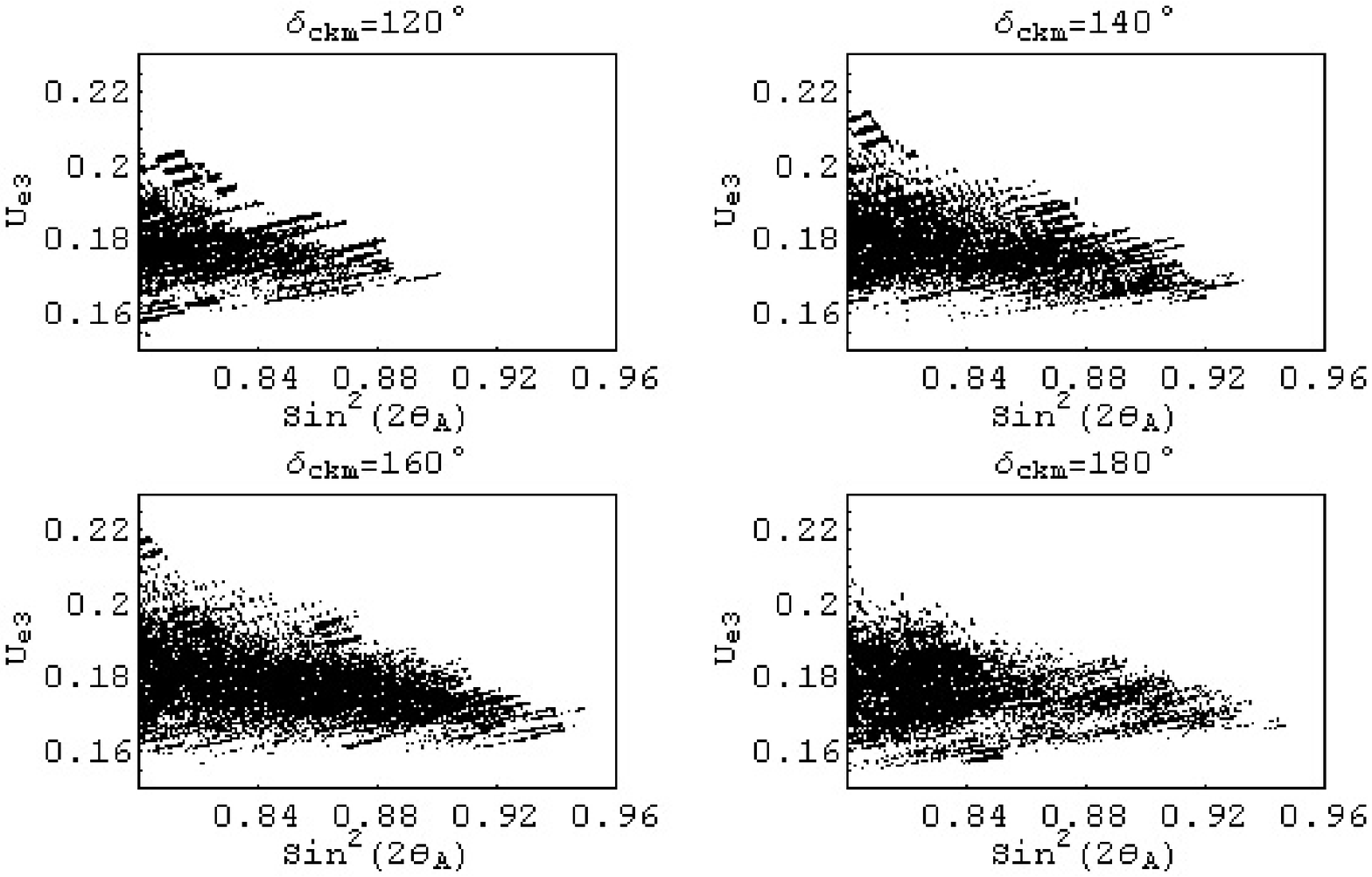}
\caption{
The figure shows the predictions for $\sin^22\theta_{A}$ and
 $U_{e3}$ for the range of quark masses and
mixings that fit charged lepton masses and where all CP phases in the
fermion masses are included, for four different values of $\delta_{CKM}$.
\label{fig:cstr6}}
\end{center}
\end{figure}

In concluding this section, we note that since the main
constraints on the model came from trying to fit the electron
mass, it is worth pointing out that one way to avoid these
constraints is to include the effects of the higher dimensional
operators (HDO) which contribute to the electron mass and is too
small to be of relevance in the discussion of other masses. Since
usually the contribution of HDO terms in this theory would
contribute terms of order
$\left(\frac{M_U}{M_{P\ell}}\right)^n\simeq 10^{-2n}$, where $3+n$
denotes the dimensionality of the HDO. The leading order HDO must
therefore correspond to $n=2$. This can be guaranteed for instance
by demanding that superpotential is R-odd and all the Higgs fields
are R-odd, whereas the matter fields are R-even. We have repeated
our computations with the HDO terms mentioned and we find that
one can obtain the neutrino mixing parameters in the right range for
the standard model value of the CKM phase.

\subsection{CP violation in leptonic mixings}
The neutrino mass matrix discussed above can be approximately diagonalized
to obtain the CP phases in the PMNS matrix. For this purpose, we first
diagonalize the charged lepton mass matrix. The mixing matrix has a form
similar to the CKM matrix. Therefore, in the basis where charged leptons
are mass eigenstates, the neutrino mass matrix remains approximately of
the same form as in Eq. (23) i.e.
\begin{eqnarray}
{\cal M}_\nu &=&m_0 zA\lambda^2\pmatrix{0 & 0 & a e^{i(\delta+\alpha_3)}
\cr
0 & (1+\epsilon)e^{i\beta_2} & e^{i\alpha_3} \cr
 a e^{i(\delta+\alpha_3)} & e^{i\alpha_3} &
\frac{(ye^{i\beta_3}-ze^{i\alpha_3})}{zA\lambda^2}}
\end{eqnarray}
where $a=\sqrt{(1\pm \rho)^2+\eta^2}~zA\lambda$; $\epsilon~\simeq
\frac{0.7y}{z}-1$ and $\tan \delta\simeq \frac{\eta}{1- \rho}$ .
In the expression for $a$ and $\tan\delta$, the plus sign
corresponds to the case without any contribution from the higher
dimension terms to the electron  mass and the minus corresponds to
the case with higher dimensional terms. In order to get maximal
mixing out of b-tau mass convergence, we must have
$\alpha_3=\beta_3$ and $(ye^{i\beta_3}-ze^{i\alpha_3})\simeq
zA\lambda^2$. Furthermore, to get the solar mass difference right,
we must have $\beta_2=\alpha_3$. We can therefore take the common
phase $e^{i\alpha_3}$ out of the matrix which leaves us with a
simple matrix of the form
\begin{eqnarray}
\tilde{{\cal M}}_\nu &=&\pmatrix{0 & 0 & a
e^{i\delta}
\cr
0 & (1+\epsilon) & 1 \cr
 a e^{i\delta} & 1 & 1}
\end{eqnarray}
diagonalize. The matrix that diagonalizes it is of the form
\begin{eqnarray}
{\bf \Large U}~=~V\left(\begin{array}{ccc} c & -s & {a\over
2}\sqrt{2}\\
-\frac{(s(1-{\epsilon\over 4})+c\sqrt{2}{a\over 4})}{\sqrt{2}} &
\frac{(-c(1-{\epsilon\over 4})+s\sqrt{2}{a\over 4})}{\sqrt{2}} &
\frac{1+{\epsilon\over 4}}{\sqrt{2}} \\
\frac{(s(1+{\epsilon\over 4})-c\sqrt{2}{a\over 4})}{\sqrt{2}} &
\frac{(c(1+{\epsilon\over 4})+s\sqrt{2}{a\over 4}) }{\sqrt{2}} &
\frac{1}{\sqrt{2}} \end{array}\right)
\end{eqnarray}
where ${\bf V} ~=~
 diag(e^{i\delta}, 1, 1)$. Note that since the PMNS mixing matrix is
real to order $\lambda$, direct CP violating effects among neutrinos are
suppressed and  unobservable in near future. This is a definite prediction
of the minimal SO(10) model. In a sense this is not surprising, since the
lepton sector and the quark sector are linked by quark-lepton unification
and we know that in the quark sector the CP violating phase is only
present in the $\lambda^3$ order.

Since the model predicts a large $U_{e3}$, if there is any CP violating
phase in the mixing matrix, it has a better chance of being observable; on
the other hand this model predicts no CP violation, so that the model has
a better chance of being falsifiable and strengthens the case for
searching for CP violation in neutrino oscillations. It is important to
emphasize that the model however enough CP violation to in the right
handed neutrino couplings so that one should expect enough baryogenesis.

\section{Summary and conclusions}
In summary, we have shown that the minimal SO(10) model with
fermion masses receiving contributions from only one {\bf 10} and
one {\bf 126} Higgs multiplets is fully predictive in the neutrino
sector. It predicts all three neutrino mixing angles in agreement
with present data but with a value of $U_{e3}$ which is very close
to the present upper limits from the reactor experiments. This
high value of $U_{e3}$ provides a test of the model. We also find
that the introduction of CP phases in the Yukawa couplings still
keeps the model predictive. The CKM phase in this case is outside the one
$\sigma$ region of the present central value in the standard model
 suggesting that there are new CP
violating contributions from the SUSY breaking sector. We find it
interesting that to the leading order in the Cabibbo angle, the
leptonic CP violation vanishes in spite of the fact that the quark
sector has CP violation.

There are no additional global symmetries assumed in the analysis. The
neutrino data once refined would therefore provide a crucial test
of minimal SUSY SO(10) in the same way as proton decay was considered a
crucial test of minimal SU(5).

\bigskip

This work is supported by the National Science Foundation
Grant No. PHY-0099544.

\end{document}